\documentclass[usenatbib,usegraphicx,letterpaper]{mn2e}
\usepackage[totalwidth=480pt,totalheight=680pt]{geometry}

\usepackage{amssymb}
\usepackage{epsfig}
\usepackage{amsmath}
\usepackage{color}

\newcommand{\apjl}{ApJL~}
\newcommand{\apjs}{ApJS~}

\newcommand{\unit}[1]{\mathrm{#1}}

\newcommand{\wprp}{w_{\mathrm{p}}(r_{\mathrm{p}})}

\newcommand{\Mpc}{\unit{Mpc}} 
\newcommand{\kpc}{\unit{kpc}}

\newcommand{\kms}{\unit{km \ s^{-1}}}

\newcommand{\hmpcvol}{h^{-3}\mathrm{Mpc^{3}}}

\newcommand{\hhMsun}{h^{-2}M_{\odot}}

\newcommand{\Mstar}{M_{\ast}}

\newcommand{\Lstar}{L^{\ast}}

\newcommand{\MBCG}{M_{\ast}^{\mathrm{BCG}}}

\newcommand{\Msun}{M_{\odot}}

\newcommand{\pimax}{\pi_\mathrm{max}}

\newcommand{\vmax}{\mathrm{V}_\mathrm{max}}

\newcommand{\vpeak}{\mathrm{V}_\mathrm{peak}}

\newcommand{\dd}{\mathrm{d}}
\newcommand{\nh}{n_{\mathrm{h}}}

\newcommand{\dndvmaxprime}{\frac{\dd\nh}{\dd\vmax'}}
\newcommand{\ngalaxy}{n_{\mathrm{g}}}
\newcommand{\nhcumulative}{N_{\mathrm{h}}(>\vmax)}
\newcommand{\ngcumulative}{N_{\mathrm{g}}(>\Mstar)}

\newcommand{\SMcat}{{M}_{\ast}^{9.8}}
\newcommand{\SMcatb}{{M}_{\ast}^{10.2}}
\newcommand{\SMcatc}{{M}_{\ast}^{10.6}}

\newcommand{\mchar}{M_{\mathrm{char}}}

\newcommand{\zform}{z_{\mathrm{form}}}
\newcommand{\zacc}{z_{\mathrm{acc}}}
\newcommand{\zchar}{z_{\mathrm{char}}}
\newcommand{\zstarve}{z_{\mathrm{starve}}}
\newcommand{\xhalo}{X_{\mathrm{halo}}}

\newcommand{\veff}{V_{\mathrm{eff}}}

\newcommand{\Psdss}{P_{\mathrm{SDSS}} ( g-r | M_{*} )}

\newcommand{\PNM}{P(N|M)}

\newcommand{\beq}{\begin{equation}}
\newcommand{\eeq}{\end{equation}}
\newcommand{\beqray}{\begin{eqnarray}}
\newcommand{\eeqray}{\end{eqnarray}}

\newcommand{\ben}{\begin{enumerate}}
\newcommand{\een}{\end{enumerate}}
\newcommand{\bit}{\begin{itemize}}
\newcommand{\eit}{\end{itemize}}

\usepackage{epsfig}  \usepackage{graphicx}   \usepackage{rotating}

\begin{document}

\title[The Dark Side of Galaxy Color]
{The Dark Side of Galaxy Color: evidence from new SDSS measurements of galaxy clustering and lensing}


\author[Hearin et al.]
{Andrew P. Hearin$^{1}$, Douglas~F.~Watson$^{2,}$\thanks{NSF Astronomy \& Astrophysics Postdoctoral Fellow},
 Matthew~R.~Becker$^{2,3,4}$,
 \newauthor
Reinabelle Reyes$^{2}$, 
Andreas~A.~Berlind$^{5}$, Andrew R.~Zentner$^{6,7}$ \\
$^1$Fermilab Center for Particle Astrophysics, Fermi National Accelerator Laboratory, Batavia, IL \\
$^2$Kavli Institute for Cosmological Physics, 5640 South
  Ellis Avenue, The University of Chicago, Chicago, IL \\
$^3$ SLAC National Accelerator Laboratory, Menlo Park, CA 94025 \\
$^4$ Kavli Institute for Particle Astrophysics and Cosmology, Stanford, CA 94309, USA \\
$^5$ Department of Physics and Astronomy, Vanderbilt University, Nashville, TN \\
 $^6$ Department of Physics and Astronomy, University of Pittsburgh, Pittsburgh, PA 15260 \\
 $^7$ Pittsburgh Particle physics, Astrophysics and Cosmology Center (PITT PACC)
}

\maketitle

\begin{abstract}

The age matching model has recently been shown to predict correctly the
luminosity $L$ and $g-r$ color of galaxies residing within dark matter
halos. The central tenet of the model is intuitive: older halos tend to host galaxies with older stellar populations. 
In this paper, we demonstrate that age matching also correctly predicts the $g-r$ color trends exhibited in a wide variety of statistics of the galaxy distribution 
for stellar mass $\Mstar$ threshold samples. In particular, we present new measurements of the galaxy two-point
correlation function and the galaxy--galaxy lensing signal
$\Delta\Sigma$ as a function of $\Mstar$ and $g-r$ color from the
Sloan Digital Sky Survey, and show that age matching exhibits remarkable agreement with these and other statistics of low-redshift galaxies. 
In so doing, we also demonstrate good agreement between the galaxy-galaxy lensing observed by SDSS and the $\Delta\Sigma$ signal predicted by abundance matching,  
 a new success of this model.
 We describe how age
 matching is a specific example of a larger class of
Conditional Abundance Matching models (CAM), a theoretical framework
we introduce here for the first time.  CAM  provides a general formalism to study correlations at fixed mass between {\em any} galaxy property and {\em any} halo property. 
The striking success of our
simple implementation of CAM provides compelling evidence that this
technique has the potential to describe the same set of data as
alternative models, but with
a dramatic reduction in the required number of parameters. 
CAM achieves this reduction by exploiting the capability of contemporary N-body simulations to determine dark matter halo properties other than mass alone, 
which distinguishes our model from 
conventional approaches to the galaxy-halo connection. 

\end{abstract}

\begin{keywords}
  cosmology: theory --- dark matter --- galaxies: halos --- galaxies:
  evolution --- galaxies: clustering --- large-scale structure of
  universe
\end{keywords}


\section{INTRODUCTION}
\label{sec:intro}


Extensive effort has been put forth to understand how the star
formation activity in galaxies maps to dark matter halos in a
cosmological context, typically via galaxy color
\citep{zehavi05a,zehavi11,skibba_sheth09,tinker_wetzel10,wang_etal07,krause_etal13,gerke12,masaki13,HW13a}.
Color is a commonly employed observable as it is correlates well with  the star formation history of a galaxy: blue galaxies exhibit
ongoing star formation, and conversely, red galaxies are typically not actively
forming stars.  Observations have long shown that redder galaxies
preferentially occupy more dense environments, while bluer galaxies
tend to reside in underdense regions
\citep{balogh99,blanton05,weinmann06b,weinmann09,peng_etal10,peng_etal12,carollo_etal12}.
Additionally, there exists a clear bimodality in the distribution of galaxy
colors, with distinct red (ellipsoidal) and blue (disky) populations
\citep{blanton03,baldry04,blanton05,wyder07}. This segregation between
 galaxy populations is already in place at $z\sim1$ \citep{bell04,cooper06,cooper12} and possibly extends out to $z\sim3$ \citep{whitaker11}.

Recently, \citet{HW13a} introduced {\em age distribution matching} (or
simply {\em age matching} for brevity), a new theoretical formalism
for connecting galaxies to halos as a function of their color and
luminosity.  Age matching is rooted in the popular abundance matching
technique, wherein the halo maximum circular velocity $\vmax$ (or
mass) is in monotonic correspondence with luminosity $L$ (or stellar
mass).  This simple, yet powerful approach has been employed to
describe a variety of observed galaxy statistics
\citep[e.g.,][]{kravtsov04a, vale_ostriker04,tasitsiomi_etal04,
  vale_ostriker06, conroy06, conroy_wechsler09,guo10, simha10,
  neistein11a, watson_etal12b, reddick12,rod_puebla12,
  hearin_etal12b}.    However, traditional abundance matching
does not  capture well-established features in the galaxy
distribution, such as color bi-modality, that reflect the complexity
in the physics of  galaxy evolution.

To that end, \citet{HW13a} (hereafter Paper I) showed that by
extending the traditional abundance matching formalism to consider an
additional halo property beyond $\vmax$, the observed spatial
distribution of galaxies as a function of luminosity {\em and} color
could be accurately reproduced.  Specifically, the authors considered
the redshift, dubbed $\zstarve$, that correlates with the epoch at
which the star formation in the galaxy is likely stifled, ultimately
leading to the quenching of the galaxy. 
By using merger trees to map the full mass assembly history (MAH) of halos, a halo's $\zstarve$ value is determined  by whichever of the following three events happens first in its MAH: (1) the epoch a halo accretes onto a larger halo, thus becoming a subhalo, (2) the epoch  a halo reaches a characteristic mass\footnote{\citet{HW13a} tried several values for  the characteristic mass and found $10^{12}\Msun$ to be most  compatible with the data.}, and (3) the epoch a halo transitioned from the fast- to slow-accretion regime.  
Under the simple assumption that $\zstarve$ correlates with $g-r$ color at fixed luminosity, the age matching technique was
able to accurately predict color-dependent clustering in the Sloan Digital Sky
Survey \citep[SDSS:][]{york00a,DR7_09} and a variety of galaxy group
statistics. The success of the model supported the idea that the assembly
history of galaxies and halos are correlated.

The central tenet of age matching is a simple one: older galaxies
live in older halos.  Of course, other halo proxies beyond $\zstarve$
can be investigated, thus age matching can be thought of
as a specific implementation of a more general formalism we will call
{\em Conditional Abundance Matching} or CAM.  As we lay out in detail
in \S~\ref{subsec:analytic}, CAM provides the framework for probing
 correlations between any  galaxy property
(e.g., color, star formation rate, morphology) and any additional halo
property.   For the purposes of this paper, we aim to test the
predictions of age matching  against a battery of new observational
measurements from SDSS: including, {\em stellar mass}-dependent
clustering and galaxy-galaxy lensing as a function of color as well as
galaxy group statistics based on an SDSS galaxy group catalog. 

The paper is laid out as follows.  In \S~\ref{sec:data} we discuss the
data and new measurements incorporated throughout this work.  In
\S~\ref{sec:simulation} we discuss the simulation, halo catalogs, and merger
trees.  In \S~\ref{sec:model} we give an overview of our general
methodology, including a description of how our SDSS--based mock
catalog is constructed as well as an analytic presentation of CAM.  A brief
discussion of how we make predictions using CAM for various galaxy
statistics is given in \S~\ref{sec:predictions}. Results are
presented in \S~\ref{sec:results}, followed by a discussion and summary in \S~\ref{sec:discussion} $\&$ \S~\ref{sec:summary}, respectively.  Throughout this work we assume a
flat $\Lambda$CDM cosmological model with $\Omega_{\mathrm{m}}=0.27,$
and a Hubble constant of $H_{0}=70$ $\mathrm{Mpc}^{-1}\mathrm{km/s}.$


\section{DATA AND MEASUREMENTS}
\label{sec:data}

\subsection{SDSS Galaxy Sample}
\label{subsec:sample}

Our baseline galaxy catalog is a volume-limited sample of galaxies
taken from the Main Galaxy Sample of Data Release 7 (\citet{DR7_09},
DR7 hereafter) of SDSS. This is the DR7 update  of the DR3 sample used
in \citet{berlind06}, to which we refer the reader for details.  The
effective volume of this subsample is $\veff=5.8\times10^{6}\hmpcvol$. 
These galaxies span the redshift range $0.02 \leq z \leq 0.068;$ the
upper redshift bound was determined by a completeness requirement in
$r$-band absolute magnitude $M_{r}  < -19,$ where $M_r$
refers to Petrosian magnitude measurements. For convenience, we refer
to this catalog as our ``Mr19'' catalog. 

Stellar masses were taken from the MPA-JHU catalog, publicly available
at {\tt http://www.mpa-garching.mpg.de/SDSS/DR7}. The stellar masses
in this catalog were estimated using the {\tt kcorrect} code of
\citet{blanton_roweis07}, assuming a Chabrier IMF
\citep{chabrier03}. We study properties of galaxy samples constructed
with stellar mass cuts at $\log_{10}(\Mstar)>[9.8,10.2,10.6],$ where stellar
masses quoted in $\hhMsun$.  
We refer to these as our
$\SMcat$ catalog, our $\SMcatb$ catalog, and our $\SMcatc$ catalog,
respectively. Each of these three stellar mass-limited samples were
constructed from the Mr19 catalog. Using a $M_r  <-18$ volume-limited catalog constructed in the same fashion as our Mr19 sample, 
we estimate that the $\SMcat$ sample is $95\%$ complete in stellar mass.

For several of the statistics explored in this paper, we have divided our galaxy samples into ``red'' and ``blue'' subsamples. 
To do so, we adopt the convention of \citet{vdBosch08}, separating the galaxies with the following stellar mass-dependent cut:
\beq
\label{eq:colorcut}
g-r = 0.76 + 0.15\left[\log_{10}(\Mstar) - 10.0\right].
\eeq

\subsection{Clustering Measurements}
\label{subsec:clustering}

We measure the projected correlation function of these stellar mass threshold and color split samples in a way similar to \citet{zehavi11}.  We first count data-data, data-random, and random-random pairs as a function of line-of-sight separation $\pi$ and projected separation $r_{\mathrm{p}}$ in 15 logarithmically spaced bins of $r_{\mathrm{p}}$ in the range $0.1-20 \ \Mpc$.  We then use the \citet{landyszalay93} estimator to evaluate the correlation function: $\xi(r_{\mathrm{p}}, \pi) = \mathrm{(DD - 2DR + RR)/RR}.$  We use a random catalog with one million points to measure DR and RR in order to ensure that Poisson errors in the random counts do not dominate the error budget.  Next we integrate along the line-of-sight to compute the projected correlation function $w_{\mathrm{p}}(r_{\mathrm{p}}).$  Specifically, we integrate out to $\pi=40 \ \Mpc$, which is sufficient to remove most of the effect of redshift distortions.  Finally, we compute errors in our measurements by jackknife resampling from 50 equal-area regions on the sky.  Our measured correlation functions and error estimates for the full sample, as well as the red/blue samples, are listed in Tables \ref{tab:wp_thresholds_all}-\ref{tab:wp_thresholds_red}.

\subsection{Galaxy Group Sample}
\label{subsec:groups}

We construct a catalog of galaxy groups from the $\SMcat$ galaxy
sample described in \S~\ref{subsec:sample} using the group-finder
introduced in \citet{berlind06}, to which we refer the interested
reader for details.  Briefly, this algorithm parses galaxies into
groups via a  friends-of-friends algorithm with different linking
lengths in the transverse and line-of-sight direction; values of the
linking lengths were chosen to optimize completeness and purity of the
group sample. The algorithm has no regard for galaxy  properties
beyond their angular positions and redshifts. We define each group's
central galaxy to be the group member with the largest stellar
mass,\footnote{See \citet{skibba11} for a discussion of the
  complicating factor that central galaxies are not always the
  brightest group members.} and satellite galaxies to be the
remaining group members.  

\subsection{Galaxy-Galaxy Lensing Measurements}
\label{subsec:gglensing}

We compute the lensing signal  in 23 logarithmic radial bins from $0.2$
to $2 \ \Mpc$ as a  weighted summation over lens-source pairs, using the
following estimator:
\begin{equation} \label{eq:dsestimator}
\Delta\Sigma(R) = \frac{\sum_{ls} w_{ls} \gamma_t^{(ls)}
  \Sigma_c^{(ls)}}{2 {\cal  R}\sum_{ls} w_{ls}},
\end{equation}
where $\gamma_{\rm t}$ is the tangential shear, and the critical
surface density $\Sigma_c^{(ls)}$ is a geometric factor,
\beq\label{eq:sigmacrit} \Sigma_c^{(ls)} = \frac{c^2}{4\pi G}
\frac{D_{\rm s}}{D_{\rm l} D_{\rm ls}(1+z_{\rm l})^2}.  \eeq Here,
$D_{\rm l}$ and $D_{\rm s}$ are angular diameter distances to the lens
and source, $D_{\rm ls}$ is the angular diameter distance between the
lens and source, and the factor of $(1+z_{\rm l})^{-2}$ arises due to
our use of comoving coordinates.  The factor of 2${\cal R}$ arises due
to our definition of ellipticity and the shear responsivity ${\cal R}$
is approximately $1 - e^2 \approx 0.87$ \citep{bernstein02}.  The
weights are assigned according to the error on the shape measurement
via \beq \label{eq:wls} w_{ls} =
\frac{(\Sigma_c^{(ls)})^{-2}}{\sigma_e^2 + \sigma_{SN}^2} \eeq where
$\sigma_e$ is the estimated shape error per component and
$\sigma_{SN}$ is the intrinsic shape noise per component, which was
determined as a function of magnitude in \citealt{mandelbaum05},
figure 3. The factor of $\Sigma_c^{-2}$ converts the shape noise in
the denominator to a noise in $\Delta\Sigma$; it down-weights pairs
that are close in redshift, so that we are weighting by the inverse
variance of $\Delta\Sigma$.

There are several additional procedures that must be done when
computing the signal (see  \citealt{mandelbaum05} for details).
First, the signal computed around random points must be subtracted
from the signal around real lenses to eliminate contributions from
systematic shear.  Second, the signal must be boosted, i.e.,
multiplied by $B(R) = n(R)/n_{\rm rand}(R)$, the ratio of the weighted
number density of sources around real lenses  relative to the weighted
number density of sources around random points  in order to account
for the dilution of the lensing signal due to sources that are
physically associated with a lens, and therefore not lensed. 

To determine errors on the lensing signal and boost factors, we divide
the survey area into 200 bootstrap subregions and generate 500
bootstrap-resampled datasets. 

The source galaxy catalogue used here was introduced and described in
\citet{reyes_etal12},  which is an improved version of the catalogue
from \citet{mandelbaum05}.  The full sample covers an area of 9243
deg$^2$, containing over 39 million galaxies from SDSS DR8.  For the
lensing calculation in this work, we used a subset of that area that
overlaps with the lens galaxy sample  (around $8\%$ of the parent
galaxy sample was in areas without source galaxy shape measurements,
so were excluded from the calculation).

The shape measurements utilise a method of PSF-correction known as
re-Gaussianization \citep{hirata_seljak03}.  Re-Gaussianization is a
method based on the use of the moments of the image and of the PSF to
correct for the effects of the PSF on the galaxy shapes. However,
unlike many other moments-based corrections, it includes corrections
for the non-Gaussianity of the galaxy profile
\citep{bernstein02,hirata_seljak03} and of the PSF (to first order in
the PSF non-Gaussianity). For more details,  we refer the reader to
\S~4 and Appendices A \& B of \citet{reyes_etal12} and
\citet{mandelbaum05}.



\section{SIMULATION AND HALO CATALOGS}
\label{sec:simulation}

The foundation of our mock galaxy catalogs is the Bolshoi $N-$body
simulation \citep{bolshoi_11}.  With a force resolution of
$\epsilon=1\,\kpc,$ the simulation solves for the evolution of
$2048^{3}$ collisionless particles in a $\Lambda$CDM cosmological
model with  $\Omega_{\mathrm{m}}=0.27$, $\Omega_{\Lambda}=0.73$,
$\Omega_{\mathrm{b}}=0.042$, $h=0.7$, $\sigma_{8}=0.82$, and
$n_{\mathrm{s}}=0.95$.  The periodic box of Bolshoi has a side length
of $250\Mpc$;  each particle has mass of $m_{\mathrm{p}} \simeq
1.9\times10^8\,\Msun$.  The simulation was run with the Adaptive
Refinement Tree Code  (ART; \citealt{kravtsovART97,
  gottloeber_klypin08}). Snapshots and halo catalogs are available at
{\tt http://www.multidark.org}. We refer the reader to
\citet{riebe_etal11} for additional information.

To construct our mocks, we use ROCKSTAR merger trees and halo
catalogs \citep{rockstar_trees,rockstar}, which are publicly available at {\tt
  http://hipacc.ucsc.edu/Bolshoi/MergerTrees.html}.  ROCKSTAR
identifies and tracks halos in phase-space, and is capable of
resolving Bolshoi halos and subhalos down to $\vmax\sim55\kms$. Halo
masses were calculated using spherical overdensities according to the
redshift-dependent virial overdensity criterion of
\citet{bryan_norman98}.



\section{MODEL}
\label{sec:model}

We use a two-phase algorithm to assign stellar mass and $g-r$ color to
our halos and subhalos in the halo catalog. Our implementation is
identical to that in Paper I, except with stellar mass $\Mstar$
replacing r-band absolute magnitude. We sketch this algorithm in
\S~\ref{subsec:sham} and \S~\ref{subsec:adm}, and refer the reader to
Paper I for further details. In \S~\ref{subsec:analytic} we describe
how the age matching technique is a special case of a more general
class of CAM models, and provide an analytical formulation of CAM.

\subsection{Stellar Mass Assignment}
\label{subsec:sham}

We use the abundance matching technique to assign stellar masses to
halos and subhalos. This technique is widely used throughout the
literature, and so we limit ourselves here to a brief sketch of the
essential features. 

A halo's maximum circular velocity is defined as $\vmax\equiv
\mathrm{Max}\left\{ \sqrt{GM(<r) / r}\right\},$ where $M(<r)$ is the
mass interior to the halo-centric distance $r.$ Abundance matching
requires that the observed cumulative abundance of galaxies as a
function of stellar mass, $N_{g}(>\Mstar)$, is equal to the cumulative
abundance of (sub)halos with circular velocities larger than $\vmax$,
$N_{h}(>\vmax)$. This requirement uniquely determines a monotonic
relationship between $\Mstar$ and $\vmax;$ we use this monotonic
relation to paint stellar masses onto every (sub)halo in the
$z=0$ Bolshoi halo catalog.

To carry out the abundance matching, we use the halo property
$\vpeak,$ the largest value of $\vmax$ the halo ever attains
throughout its assembly history \citep{reddick12,behroozi13c}. We
model stochasticity between stellar mass and $\vpeak$ in the exact
same fashion described in detail in Appendix A of
\citet{hearin_etal12b}, which results in uniform scatter in stellar
mass of $\sim 0.15$dex at fixed $\vpeak,$ a level of scatter that is
consistent with that found in a variety of other studies
\citep{more09b,reddick12,hearin_etal12b}.

\subsection{Color Assignment}
\label{subsec:adm}

After assigning stellar masses to mock galaxies, we proceed to  the
second phase of our algorithm, in which we assign $g-r$ colors. First,
we bin the mock and SDSS galaxies by stellar mass, using ten
logarithmically spaced bins spanning the range of both
samples.\footnote{We have performed a variety of explicit tests to
  verify that our results are not sensitive to our bin width.}  The
$g-r$ colors of the SDSS  galaxies in each stellar mass bin
empirically define the probability distribution $\Psdss.$ For the  $N$
mock galaxies in each stellar mass bin, we randomly draw $N$ times
from $\Psdss,$  rank-ordering each bin's draws, reddest first. We
assign these colors to the mock galaxies in the corresponding stellar
mass bin after first rank-ordering these $N$ mock galaxies by the
property $\zstarve$ (defined below), largest first. 

After carrying out  the above procedure in each stellar mass bin, the
$g-r$ distribution of the mock galaxies is in exact agreement with the
data, by construction. This agreement is illustrated in the top left
and bottom panels of Fig.~\ref{fig:sm_color_PDFs}.  Note that the
rank-ordering has no impact on the agreement between the observed and
mock color PDFs. The only effect of the rank-ordering is to introduce,
at fixed stellar mass, a correlation between galaxy color and the
epoch in a halo's MAH presumed to be linked to the stifling of star
formation, $\zstarve$.

Three characteristic epochs in the main progenitor history of a halo
determine its $\zstarve$ value:  \ben
\item[\textbf{1.}]{ $\zchar:$ The first epoch at which halo mass
  exceeds a characteristic mass of $M_{\mathrm{char}}=10^{12}\Msun.$
  For halos that never attain this mass, $\zchar=0.$}
\item[\textbf{2.}]{$\zacc:$  For subhalos, this is the epoch after
  which the object always remains a subhalo. For host halos,
  $\zacc=0.$}
\item[\textbf{3}]{$\zform:$ We follow \citet{wechsler02} in our
  definition of the formation epoch of a halo, using their halo
  concentration-based proxy for halo age\footnote{We use the $z=0$ concentration for host halos, and the concentration at the time of infall 
  for subhalos.}. This  identifies  the
  redshift at which the halo transitions from the fast- to
  slow-accretion regime, and the circular velocity $\vmax$ of the halo
  plateaus.}   \een  After computing each of these three
  characteristic epochs, we define the redshift of  starvation:  \beq
\label{eq:zstarvedef}
\zstarve\equiv \mathrm{Max}\left\{\zacc,\zchar,\zform\right\}.   \eeq
For details concerning how exactly to calculate $\zstarve$ from halo
merger trees, we refer to the appendix of Paper I. 

\subsection{Analytical Formulation of CAM}
\label{subsec:analytic}

Age matching admits a relatively simple analytical description that
makes clear why we consider our model to be a specific case of a
larger class of CAM models. We first
consider the analytical formulation of traditional abundance matching
to highlight this connection.

The basic outcome of the abundance matching prescription is to
determine $P(\Mstar|\vmax),$ the probability that a (sub)halo with
circular velocity $\vmax$ hosts a galaxy with stellar mass $\Mstar.$ 
In the absence of scatter between $\vmax$ and $\Mstar,$\footnote{See \citet{behroozi10} for a
  discussion of the analytical formulation of abundance matching in
  the presence of scatter.} abundance matching treats
$P(\Mstar|\vmax)$ as a delta function centered at the value
$\Mstar(\vmax),$ the mean circular velocity of the halo of a galaxy with stellar mass $\Mstar.$ 
To see how the map $\Mstar(\vmax)$ is determined analytically, let $\dd\nh/\dd\vmax$
denote the abundance of dark matter (sub)halos as a function of
$\vmax,$ and $\dd\ngalaxy/\dd\Mstar$ the observed stellar mass
function (SMF). Then the cumulative abundances are given by  \beqray
\label{eq:cumulative}
\nhcumulative &=& \int_{\vmax}^{\infty}\dd\vmax^{\prime}\dndvmaxprime
\\ \ngcumulative &=&
\int_{\Mstar}^{\infty}\dd\Mstar^{\prime}\frac{\dd\ngalaxy}{\dd\Mstar^{\prime}}
\eeqray

In the absence of scatter, traditional
abundance matching assumes that $\Mstar$ and $\vmax$ are in perfect
monotonic correspondence in such a way that the cumulative abundances
agree at all stellar masses. For any value of $\vmax,$ this can be
accomplished analytically by simply finding the zero of the following
function

\beqray
\label{eq:sham}
{F}_{\vmax}(\Mstar) &\equiv & \nhcumulative - \ngcumulative.  \eeqray
Equation \ref{eq:sham} determines the map $\Mstar(\vmax),$ as well as the associated Jacobian ${\dd\Mstar}/{\dd\vmax}$ and its inverse. 

In CAM, we seek to specify a more complicated probability distribution function (PDF),
$P(\Mstar,c|\vmax,\xhalo),$ the probability that a galaxy of stellar mass $\Mstar$ and color $c$ resides in a halo with
circular velocity $\vmax$ and an additional halo property $\xhalo$, where $\xhalo = \zstarve$ in the case of age matching.
We begin by
making the following assumption of separability: \beq
\label{eq:separability}
P(\Mstar,c|\vmax,\xhalo) = P(\Mstar|\vmax) \times
P(c|\vmax,\xhalo).  \eeq

We compute the first factor on the right hand side of
Eq.~\ref{eq:separability} by using traditional abundance matching to
determine $\Mstar(\vmax)$ (and hence $P(\Mstar|\vmax)$), so that we
may piggyback on the well-known successes of this simple approach to
galaxy-halo modeling. However, this is not a necessary feature of age
matching; one may instead wish to use, for example, a conditional
stellar mass function approach to paint stellar masses onto halos and subhalos. 

The second factor on the right hand side of Eq.~\ref{eq:separability}
is related to $P(c|\vmax)$ through simple parameter marginalization:
\beq
\label{eq:marginalization}
P(c|\vmax) = \int\dd\xhalo P(c|\vmax,\xhalo)P(\xhalo|\vmax),
\eeq where $P(\xhalo|\vmax)$ is tabulated directly from the
simulation. 

We relate the left hand side of Eq.~\ref{eq:marginalization} to the observed 
color distribution:
\beq
\label{eq:cam}
P(c|\vmax(\Mstar))\frac{\dd\vmax}{\dd\Mstar} = P_{\mathrm{DATA}}(c|\Mstar).  
\eeq 
Eq.~\ref{eq:cam} guarantees that the color PDF of the model will be in exact agreement with the data at all stellar masses, in direct analogy 
to the way  traditional abundance matching guarantees that the model will have the correct stellar mass function. 
Putting the above pieces together, we have 
\beqray
\label{eq:cam2}
P_{\mathrm{DATA}}(c|\Mstar)\frac{\dd\Mstar}{\dd\vmax} &=& \int\dd\xhalo
P(\xhalo|\vmax) \\ \nonumber
&\times& P(c|\vmax,\xhalo).  
\eeqray 
The quantity
$P(c|\vmax,\xhalo)$ is the fundamental quantity we wish to
determine. We do so in an exactly analogous fashion to standard
abundance matching.

First we define the following two conditional cumulative abundances:
\beqray
\label{eq:conditionalabun}
N_{\mathrm{h}}(>\xhalo|\vmax) &=&
\int_{\xhalo}^{\infty}\dd\xhalo^{\prime}\Bigg(\frac{\dd\nh}{\dd\xhalo^{\prime}}\Bigg)|_{\vmax} \nonumber \\
N_{\mathrm{g}}(>c|\Mstar) &=& \int_{c}^{\infty}\dd c^{\prime}
\Bigg(\frac{\dd\ngalaxy}{\dd c^{\prime}}\Bigg)|_{\Mstar}, 
\eeqray where $({\dd\nh}/{\dd\xhalo^{\prime}})|_{\vmax}$ is the abundance of
halos with circular velocity $\vmax$ as a function of $\xhalo,$ and
$({\dd\ngalaxy}/{\dd c^{\prime}})|_{\Mstar}$ is the abundance of
galaxies with stellar mass $\Mstar$ as a function of color. In the
absence of scatter, $P(c|\vmax,\xhalo)$ is simply a delta function
centered at the map $c(\vmax,\xhalo),$ which is computed by finding
the zero of 
\beqray
F_{\vmax,\xhalo}(c) &\equiv & N_{\mathrm{h}}(>\xhalo|\vmax) \\ \nonumber
&-& N_{\mathrm{g}}(>c|\Mstar).
\eeqray

The technique described above is very general; in principle it can be
applied to any observable property of galaxies, and can be implemented
with any halo property. Since the model we study in this paper uses
$\xhalo=\zstarve$ as the second halo property, which is primarily driven by
formation time, we refer to this particular implementation as age
matching. We discuss other possible implementations and applications
of the CAM formalism in \S~\ref{sec:discussion}. 



\section{PREDICTIONS}
\label{sec:predictions}

We evaluate the success of our model by computing statistics of our
mock galaxy sample in a directly analogous fashion to the manner in
which the corresponding statistics are computed in observational data.
We describe our technique for computing projected clustering of mock
galaxies in \S~\ref{subsec:predicting_wp}, the mock galaxy-galaxy
lensing in  \S~\ref{subsec:predicting_gglensing}, and mock galaxy
group statistics in \S~\ref{subsec:predicting_groups}. 

\subsection{Projected Clustering of Mock Galaxies}
\label{subsec:predicting_wp}

The projected clustering of galaxies $w_{\mathrm{p}}(r_{\mathrm{p}})$ quantifies 
the probability in excess of random that a pair of galaxies in a sample will have line-of-sight separation less than
$\pi_{\mathrm{max}}$ and be found at projected separation
$r_{\mathrm{p}}.$ Suppose that $N_{\mathrm{g}}$ galaxies are randomly
distributed throughout a cubical box of side length
$L_{\mathrm{box}}.$  Then in a randomly placed cylindrical annulus of
length $\pi_{\mathrm{max}},$ and with inner and outer radii
$r_{\mathrm{min}}$ and $r_{\mathrm{max}},$ respectively, the expected
number of galaxy pairs
is $$N_{\mathrm{p,ran}}=\frac{1}{2}N_{\mathrm{g}}(N_{\mathrm{g}}-1)(\pi
r^{2}_{\mathrm{max}}-\pi
r^{2}_{\mathrm{min}})\pi_{\mathrm{max}}/L_{\mathrm{box}}^{3}.$$ By
directly computing $N_{\mathrm{p}},$ the actual number of galaxy pairs
in our mock that satisfy this line-of-sight and projected distance
criterion,\footnote{We employ the {\em distant observer
    approximation}, using the $z-$direction in the simulation to
  compute line-of-sight distances, and the $x-y$ plane to compute
  projected distances.}  we compute $1+ w_{\mathrm{p}}(r_{\mathrm{p}})
= N_{\mathrm{p}}/N_{\mathrm{p,ran}}.$ We use $\pimax=40$
$\mathrm{Mpc}$ to be consistent with the measurements made for
our observational samples.\footnote{In computing the actual pair counts in the simulation, we first place the galaxies into redshift space, though 
we find that this has a negligible effect on the projected correlation function.} We estimate errors on $w_{\mathrm{p}}$ by jackknifing the
octants of the simulation box.

\subsection{Mock Galaxy-Galaxy Lensing}
\label{subsec:predicting_gglensing}

We compute the mock galaxy-galaxy lensing signal $\Delta\Sigma$ from
the particle data as follows. For every galaxy in our mock, we
compute the two-dimensional projected mass density profile in 25
logarithmic bins in radius from 0.2 to 2.0 $\Mpc$. The density from 0
to 0.2 $\Mpc$ is recorded as well for computing $\Delta\Sigma$
later. The two-dimensional mass profile is computed from a projection
over a length of 100 $\Mpc$ along the z-axis of the simulation box
(i.e., from 50 $\Mpc$ behind the galaxy to 50 $\Mpc$ in front of the
galaxy). With the two-dimensional density profiles, we then compute
$\Delta\Sigma$ around each galaxy for the $i^{\mathrm{th}}$ bin as
\begin{equation}
\Delta\Sigma_{i}=\frac{1}{\pi R_{i-1,\mathrm{max}}^{2}}\sum_{n=0}^{i-1}\Sigma_{n}A_{n}-\Sigma_{i}\ ,
\end{equation}
where $\Sigma_{i}$ is the surface density of mass in the $i^{\mathrm{th}}$ bin and 
$A_{i}=\pi(R_{i,\mathrm{max}}^{2}-R_{i,\mathrm{min}}^{2})$ is the area of each annulus. Here 
$R_{\mathrm{min},i}$ and $R_{\mathrm{max},i}$ are the minimum and maximum radius of the 
$i^{\mathrm{th}}$ annulus.  The errors on $\Delta\Sigma$ are computed via 27 jackknife regions 
over the simulation volume. 

The procedure for computing $\Delta\Sigma$ defined above is
approximate for several reasons. First, in the limit that each galaxy
is in a thin lens at the galaxy's redshift in both the simulation and
the data, the procedure defined above is an exact match to how
$\Delta\Sigma$ is computed from observational data. However, in
reality, a given galaxy is not in a thin lens, but is instead in an
extended mass distribution. Thus along the projected line-of-sight, the lensing kernel will supply an additional
line-of-sight dependent weight. The exact form of this weight depends
in detail on the redshift distributions of the lenses and sources in
the observational samples. Given that the galaxy-matter correlation
function falls off quickly as a function of radius, we have chosen to
 neglect this effect. Second, the projection along the
line-of-sight should be done over a full light cone. However, again
because the galaxy-matter correlation function falls off quickly as a
function of radius, we have fixed the line-of-sight projection length
to 100 $\Mpc$ (see also \citealt{leauthaud11b}). Finally, note also
that the box length for Bolshoi is only 250 $\Mpc$ and that the
simulation volume is periodic. Thus particles in the volume can be
separated by at most half the box length, 125 $\Mpc$, and so we can
project the mass distribution only over at most this length. Thus our
chosen projection length is a compromise between the need to make it
as large as possible in order to match the data accurately and the
constraints of the given simulation volume.

\subsection{Group Identification of Mock Galaxies}
\label{subsec:predicting_groups}

To parse the $\SMcat$ mock galaxies into groups, we use the
simulation $z-$coordinate together with the peculiar velocity in the $z-$direction to place mock galaxies into redshift-space, convert $x-y$
coordinates into RA and DEC, and then apply the same group-finding
algorithm on the mock galaxies as we applied on the SDSS galaxy
sample. In this way, our mock galaxy group sample is subject to the
same systematic errors as our SDSS groups. For further details
concerning group-finding in a mock, we refer the reader to
\citet{hearin_etal12b}.



\section{Results}
\label{sec:results}

In this section, we present our main results.  In
\S~\ref{subsec:cen_sats} we show that our age matching model predicts the correct relative colors of central and satellite
galaxies.  Then in \S~\ref{subsec:2PCF_results} we demonstrate the
accuracy of our base $\SMcat$ SDSS mock catalog at reproducing new
SDSS measurements of the projected galaxy 2PCF as a function of
stellar mass, explicitly showing that our mock catalog naturally
inherits the successes of traditional abundance matching. We then compare the age matching predictions to SDSS measurements of the 2PCF split into red and blue samples.  We also investigate the success of our age matching model with new SDSS
galaxy-galaxy lensing measurements in \S~\ref{subsec:lensing_results},
a statistic which was not explored in Paper I.  As is done for
clustering, we consider the lensing signal, $\Delta\Sigma$, as a
function of stellar mass, and then for distinct red and blue subsamples.
We also test the success of our model against statistics measured from
the $\SMcat$ SDSS group catalog in \S~\ref{subsec:group_results}.
Finally, in \S~\ref{subsec:zstarve_contribution} we dissect $\zstarve$
to examine exactly how halo mass assembly is linked to shaping the colors
of galaxies within age matching.  This includes a discussion of
alternative models that we tested, and the power of the more
general CAM formalism.


\begin{figure*}
\begin{center}
\includegraphics[width=1.\textwidth]{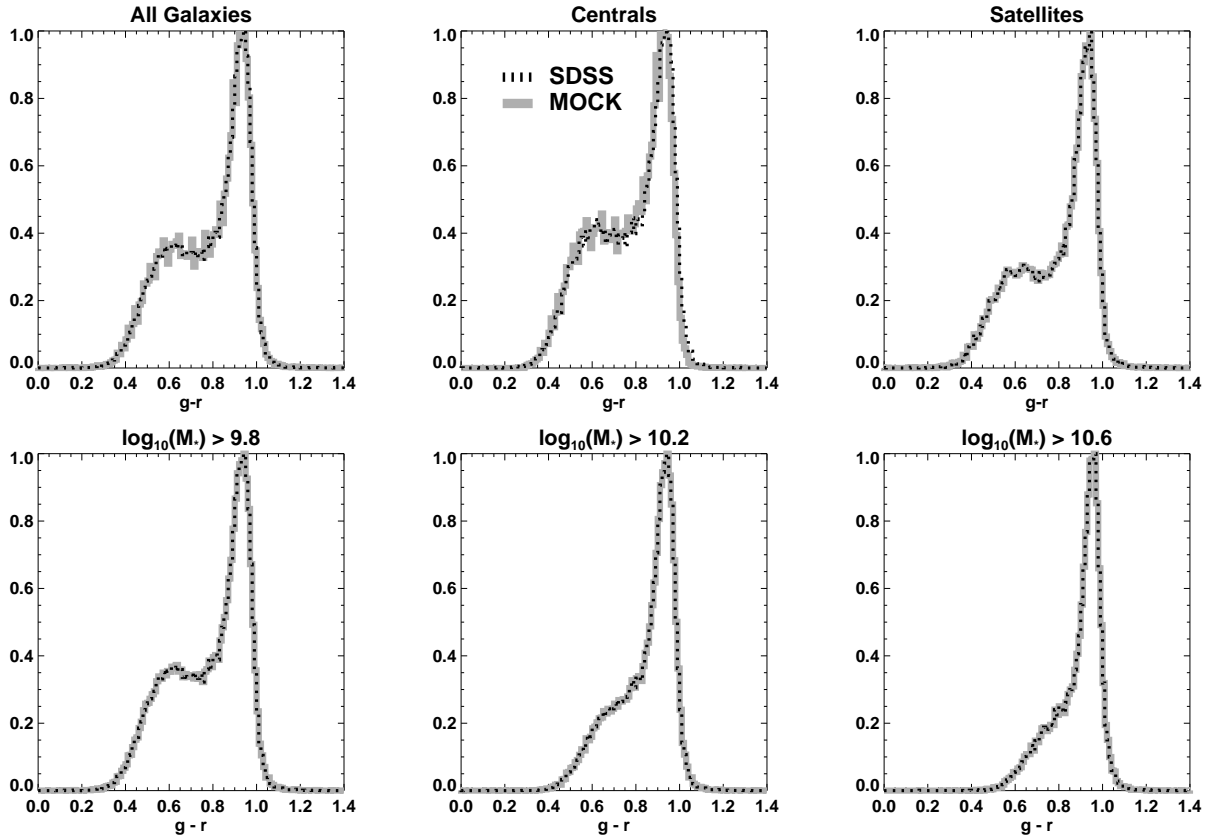}
\caption{Galaxy color probability distribution functions (PDFs) from
  our mock catalog as compared to those measured in the SDSS
  galaxy catalogs.  The $g -r $ color PDFs of our mock galaxies are in
  exact agreement with the data (black dotted histograms) for the all
  galaxy sample (top left panel) and all three stellar mass threshold
  samples (bottom row), by construction.  Our color assignment to galaxies is
  blind to central/satellite designation, thus the PDFs measured from
  the $\SMcat$ group catalog, as seen in the center and right panels of the top
  row, demonstrate the highly successful {\em predictions} of age 
  matching.}
\label{fig:sm_color_PDFs}
\end{center}
\end{figure*}



\begin{figure*}
\begin{center}
\includegraphics[width=1.\textwidth]{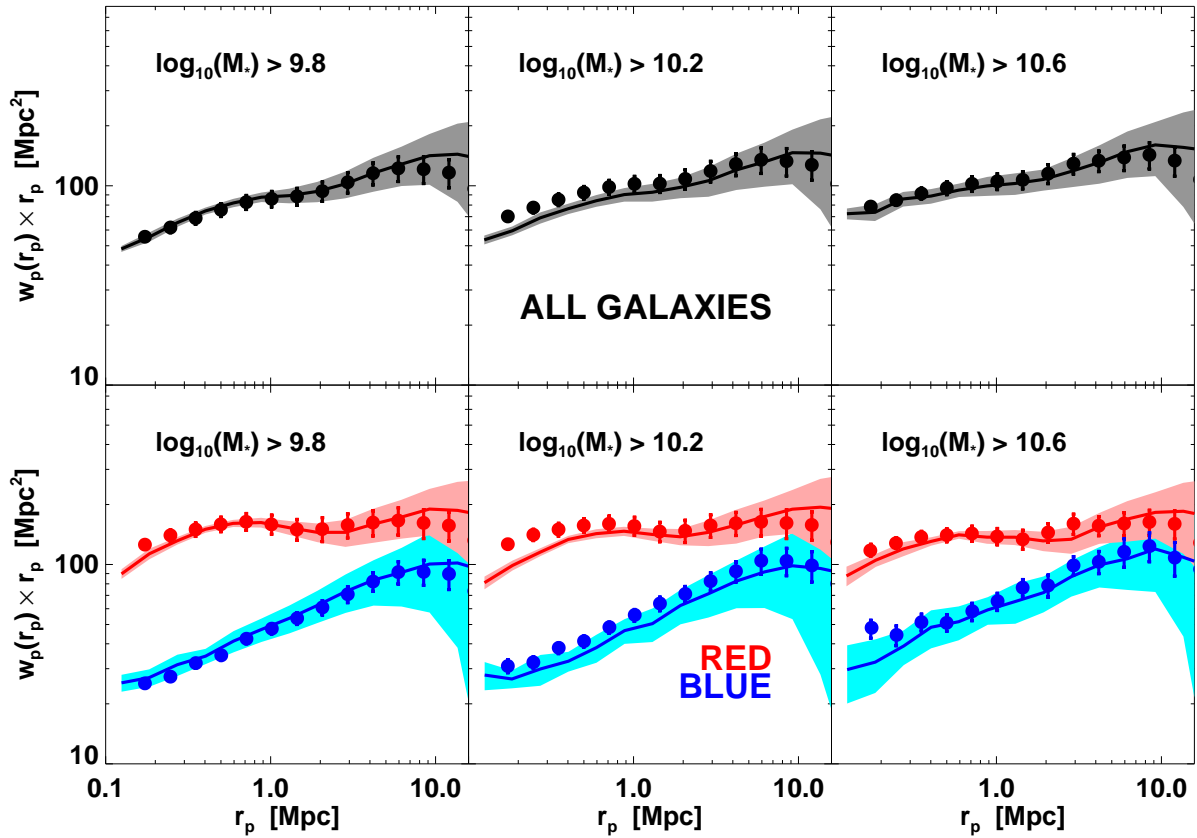}
\caption{Stellar mass- and color-dependent clustering as predicted by
  our age matching formalism. \emph{Top Row}: The projected correlation function
  (multiplied by $r_\mathrm{p}$) predicted by our model (black solid curves) as
  compared to the clustering of three SDSS stellar mass threshold samples:
  $\mathrm{log}_{10}(\Mstar) > [9.8, 10.2, 10.6]$.  \emph{Bottom
    Row}: Correlation functions split by color for red (blue) mock
  galaxies shown with red (blue) solid curves.  Red (blue) points show the clustering of red (blue) SDSS galaxies. Solid
  bands in each panel show the error in our model prediction as
  described \S~\ref{subsec:predicting_wp}.  The slight
  under-prediction of abundance matching on small scales for the
  $\mathrm{log}_{10}(\Mstar) > 10.2$ sample (top, center panel) propagates through to
  the color split (bottom, center panel), though the relative clustering strength of red and blue galaxies 
   is captured by the model at all stellar masses and over all scales.}
\label{fig:sm_color_wp}
\end{center}
\end{figure*}



\begin{figure*}
\begin{center}
\includegraphics[width=1.\textwidth]{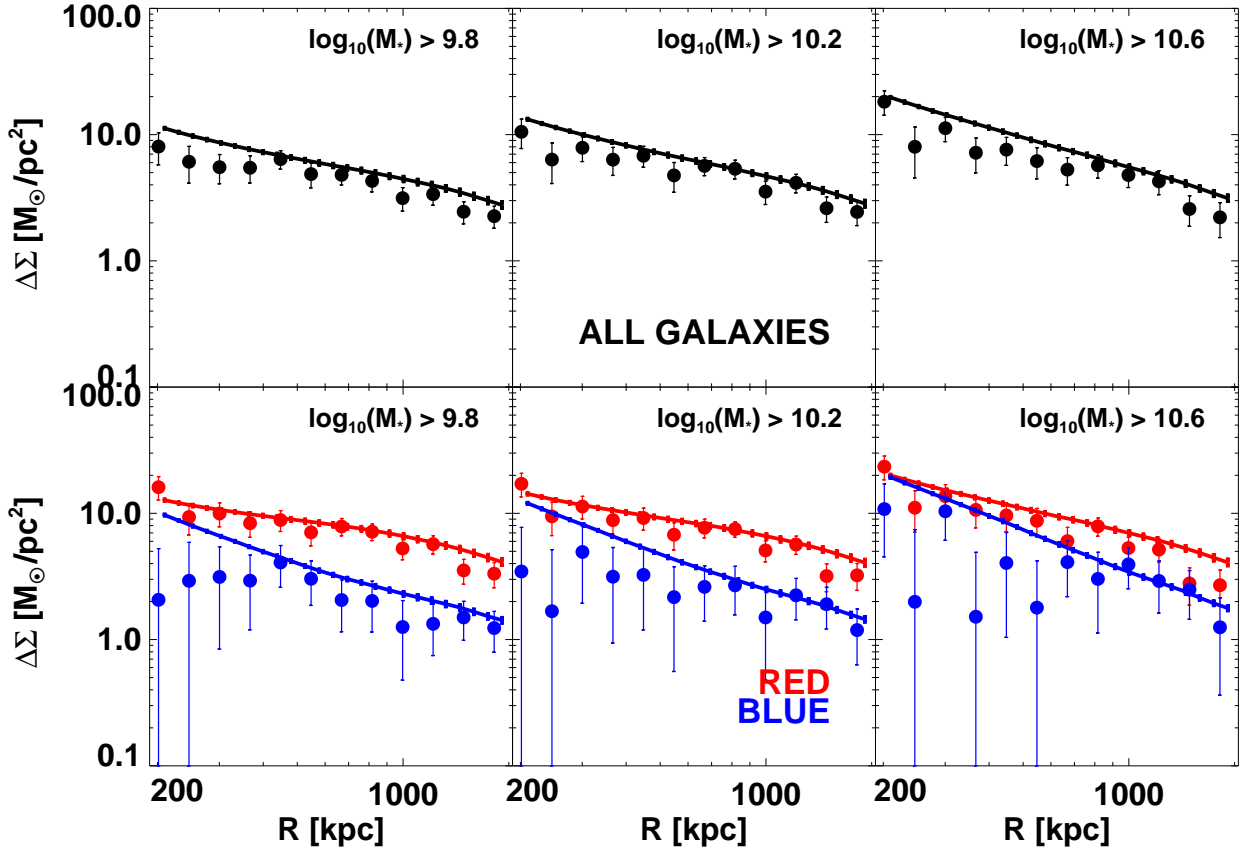}
\caption{$\Delta\Sigma$ as a function of stellar mass and color. As
  was shown for the projected clustering in
  Fig.~\ref{fig:sm_color_wp}, the top row is the abundance matching
  result for all galaxies and the bottom row shows the predicted color
  split from our age matching model (red and blue solid curves) as
  compared to new SDSS galaxy-galaxy lensing measurements.  Derived
  errors for the data and the model are described in
  \S~\ref{subsec:gglensing} $\&$ \S~\ref{subsec:predicting_gglensing},
  respectively.}
\label{fig:lensing}
\end{center}
\end{figure*}



\begin{figure*}
\begin{center}
\includegraphics[width=1.\textwidth]{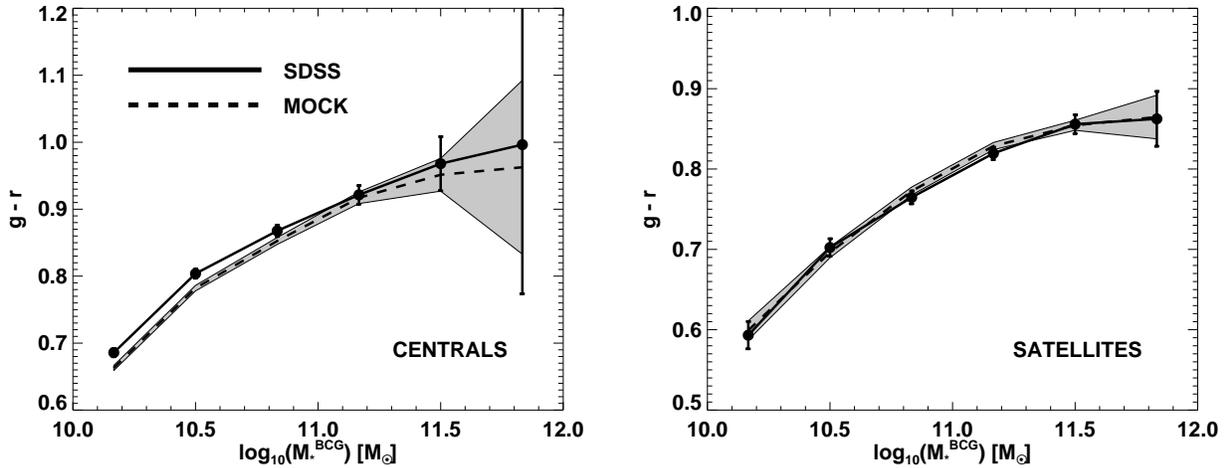}
\caption{Mean {\em g-r} color as a function of stellar mass of the
  brightest central galaxy, $\MBCG$, for SDSS galaxies (black filled
  circles) and the prediction from our mock catalog (dashed curves).
  Solid gray bands indicate Poisson error estimations.  Results for central
  galaxies are shown in the left panel and satellite galaxies in the
  right. }
\label{fig:groups}
\end{center}
\end{figure*}



\begin{figure*}
\begin{center}
\includegraphics[width=1.\textwidth]{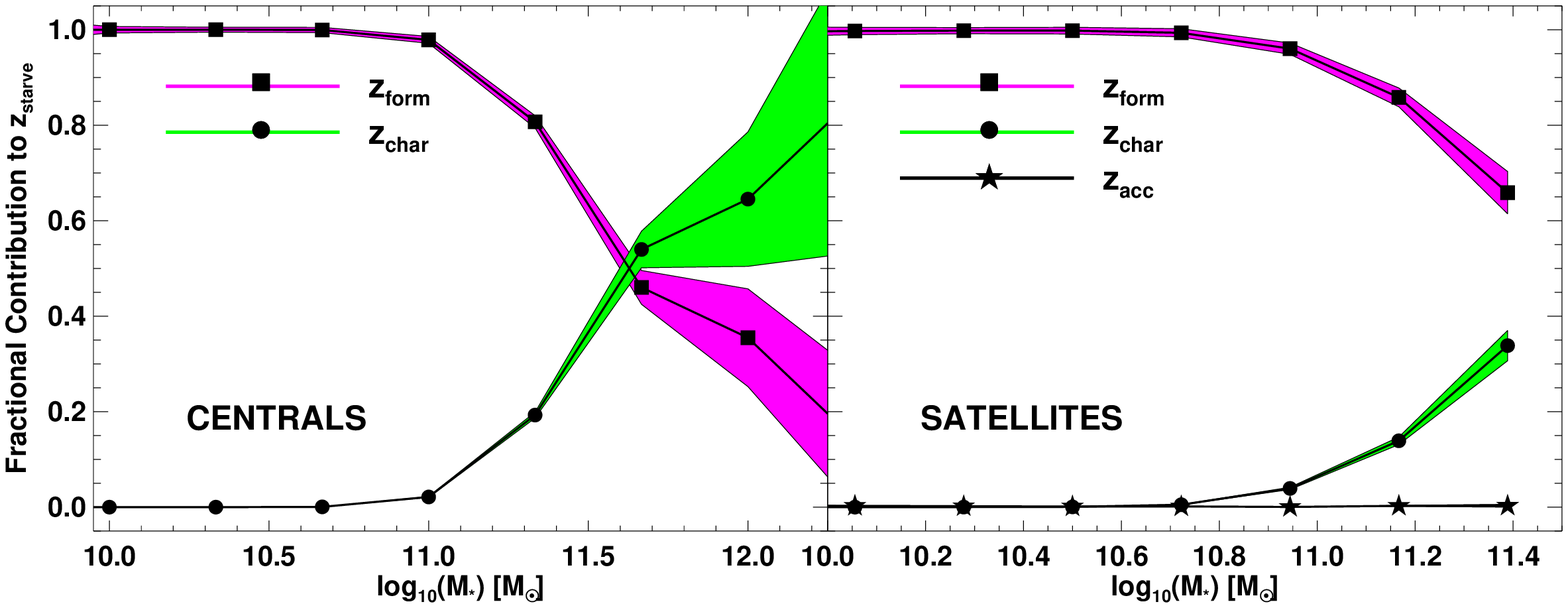}
\caption{The fractional contribution to $\zstarve$ from the three
  characteristic epochs in the mass accretion history of halos that
  constitute age matching. Results for central and satellite galaxies
  are shown in the left and right panels, respectively.  As stellar mass increases, 
  $\zchar$ plays an increasingly important role in the $\zstarve$ value of a halo. On the other hand, 
  $\zacc$ makes an essentially negligible contribution at all stellar masses.}
\label{fig:zstarve_frac}
\end{center}
\end{figure*}


\subsection{Central and Satellite Colors}
\label{subsec:cen_sats}

As discussed in \S~\ref{subsec:adm}, our model correctly reproduces
the color distribution $P_{\mathrm{SDSS}}(c|\Mstar)$ by construction
as seen in the top left panel and bottom row of
Fig.~\ref{fig:sm_color_PDFs}.  However, the information about
satellite/central designation does not inform the colors we assign to
the galaxies.  Specifically, our color assignment only uses the
property $\zstarve$ and $P_{\mathrm{SDSS}}(c|\Mstar)$ to assign colors
to the mock galaxies, but does not distinguish between centrals and
satellites.  Thus, by no means is it guaranteed that our color PDFs
will be correctly predicted when conditioned on some other galaxy
property beyond stellar mass.  The top middle and right panels of
Fig.~\ref{fig:sm_color_PDFs} clearly demonstrate the successful
prediction of age matching for the separate color PDFs of  central and
satellite galaxies.  In our model, {\em central and satellite galaxies of the same stellar mass 
  have different color distributions simply because host halos and
  subhalos have different MAHs}.

\subsection{Galaxy Clustering}
\label{subsec:2PCF_results}

In the top row of Fig.~\ref{fig:sm_color_wp},  black solid curves with
gray error bands (see section \S~\ref{subsec:predicting_wp} for a
discussion of error estimates) show the projected 2PCF measured from
our mock catalog for all galaxies predicated on traditional abundance matching.
The  agreement with the SDSS data points (filled black
circles) illustrates that galaxies have been properly assigned to
halos as a function of stellar mass.  However, notice there is a
slight under-prediction from abundance matching on small scales for
the $\SMcatb$ sample.

We now turn to the bottom row of Fig.~\ref{fig:sm_color_wp} to
investigate the success of age matching at predicting color-dependent
clustering.  Red and blue filled circles in all panels represent the
red and blue galaxy populations from SDSS, respectively.  Red and blue
solid curves are the age matching model predictions.    The
$\SMcat$ and  $\SMcatc$ predictions for the color-dependent clustering are in excellent
agreement with the data at all scales.  However, as noted above, there
is a slight under-prediction of abundance matching on small scales
($r_\mathrm{p} \lesssim 500 \kpc$) for the $\SMcatb$ sample and this propagates through to the color split (bottom,
center panel), though the relative color split of the model agrees well with what is dictated by the data.

We emphasize that our age matching model has required no
parameter fitting to achieve the agreement between the predicted and
measured color-dependent clustering. Our algorithm for color
assignment has no explicit dependence on halo position; the clustering
signal in our mock simply emerges as a {\em prediction} of age
matching.  The success of our model is compelling given the simplicity
of age matching, and the more general CAM formalism.

\subsection{Galaxy-Galaxy Lensing}
\label{subsec:lensing_results}

While the 2PCF encodes rich information about the galaxy-halo connection, 
measurements of galaxy-galaxy lensing have been shown to break degeneracies between galaxy-halo parameters
 that are present when model constraints are derived from clustering measurements alone \citep[e.g.,][]{more_etal13}. 
To that end, in
Fig.~\ref{fig:lensing} we compare our model prediction to new
measurements of the stellar mass- and color-dependent galaxy-galaxy
lensing signal, $\Delta\Sigma$. As was the case for the 2PCF
comparison, we accurately predict $\Delta\Sigma$ at the abundance
matching level (black solid curves versus SDSS solid black data points
in the top row), though the amplitude of the model prediction appears
slightly boosted relative to the data for all three stellar mass
thresholds. Red and blue
filled circles in all panels represent the red and blue SDSS galaxy
populations, respectively, while red and blue solid curves are the
model predictions according to age matching. The separation in $\Delta\Sigma$ between red and blue samples is predicted 
reasonably well, excepting only blue samples on small scales, where measurement errors become large.  

\subsection{Galaxy Group Environment}
\label{subsec:group_results}

In addition to $\wprp$ and $\Delta\Sigma$, we employ a group-finder to
test how well our model predicts the scaling of central and satellite
color with host halo mass. As our proxy for halo mass we use
$\MBCG,$ the stellar mass of the group's central galaxy.  In
Fig.~\ref{fig:groups}, we show the mean  $g - r$ color of group
galaxies as a function of $\MBCG$.   We show the results for central
galaxies and satellites from left to right, respectively.  The dashed
line is the mean $g - r$ color of the mock galaxies in a given $\MBCG$
bin. The solid gray region shows Poisson errors on the mean color in
each bin. 

Our predicted mean satellite  color is in good agreement the data over the full host halo mass range probed by our galaxy sample.   
This is also true for central galaxies, excepting some
slight tension at the low $\MBCG$ end. Again, we emphasize that we
have not tuned any parameters in our model. The successful prediction
for central and satellite colors naturally emerges from the age
distribution matching formalism. Specifically, at fixed stellar mass,
the colors of our mock galaxies are drawn from the same color PDF,
$P_{\mathrm{SDSS}}(c|\Mstar)$, regardless of subhalo or host halo
designation. Moreover, our color assignment algorithm takes no
explicit account of subhalo or host halo mass. Therefore, our model's correct prediction for 
the environmental-dependence of satellite and central galaxy colors arises purely due to the 
environmental dependence of $\vmax$ and $\zstarve$ of dark matter halos and subhalos.

\subsection{Fractional Contributions to $\zstarve$}
\label{subsec:zstarve_contribution}

As laid out in \S~\ref{subsec:adm}, a halo's $\zstarve$ value is
determined by whichever of the following three events happens first in
its MAH: (1) the epoch a halo accretes onto a larger halo, $\zacc$
(this of course only pertains to satellite galaxies), (2) the epoch a
halo reaches a characteristic mass of $10^{12}\Msun$, $\zchar$, and
(3) the epoch a halo transitions from the fast- to slow-accretion
regime, $\zform$.  Figure~\ref{fig:zstarve_frac} shows the fractional
contribution to $\zstarve$ from $\zform$ (star symbols), $\zchar$
(filled circles) and $\zacc$ (triangles) as a function of galaxy
stellar mass.  Gray bands are the Poisson errors for a given stellar
mass bin. 

First consider central galaxies in the left panel; $\zform$ (solid
magenta curve) dominates the fractional contribution at $\sim100\%$ up
to $\mathrm{log}_{10}(\Mstar) = 11.0$ (roughly $\Lstar$ galaxies), and
preciptiously declines to $\sim50\%$ at  $\mathrm{log}_{10}(\Mstar) =
11.5$. The halo property $\zchar$ was originally introduced into age
matching because older halos receive redder colors, and so without
$\zchar$, high mass galaxies (e.g., BCGs residing in clusters) would
be assigned colors that are too blue since their halos are still
forming today.  As discussed in Paper I, our phenomenological model
forms no particular hypothesis for the particular physical
mechanism(s) that influence star formation within massive halos. We
simply posit that there exists a characteristic halo mass above which
star formation becomes inefficient. Physically, this presumption is
well-motivated by  hydrodynamical simulation results that
implement  AGN feedback, which can have a dramatic effect on star
formation \citep{shankar_etal06,cattaneo_etal06,teyssier11,martizzi_etal12}.\footnote{See also
  \citet{fang_etal13} for an observational investigation of
  bulge-driven quenching.} We have considered a form of the model in
which $\zchar$ is neglected, and for stellar mass-based galaxy samples
we find that the effects on all of the galaxy statistics are minimal.
We also considered a larger 2PCF stellar mass threshold
$\mathrm{log}_{10}(\Mstar) > 11.0$, and while the color split became
more pronounced with the inclusion of $\zchar$ in the model, sample
variance errors in Bolshoi  were too large to quantitatively
distinguish between models with and without $\zchar$.  However, we
note that $\zchar$ was crucial for accurately predicting the brightest
2PCF luminosity bin in Paper I. 

We see similar trends to the above for the satellite galaxies plotted
in the right panel.  Due to the rarity of massive objects, combined with the
effects of tidal mass loss, satellite galaxies rarely achieve halo
masses large enough for $\zchar$ to make an appreciable contribution
to $\zstarve$ ($\sim30\%$ at maximum).  

\begin{figure}
\begin{center}
\includegraphics[width=0.5\textwidth]{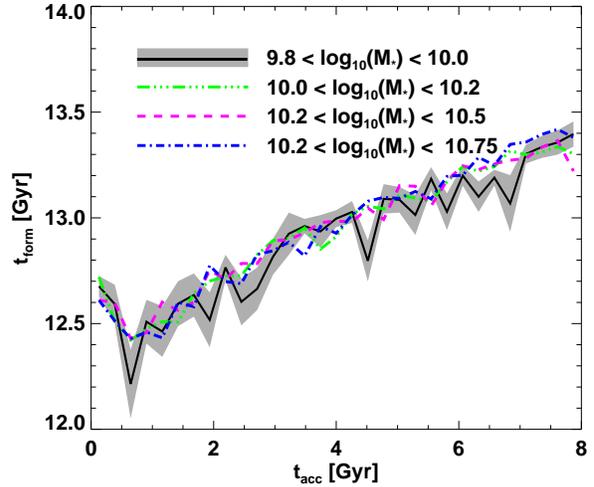}
\caption{Plot of the lookback time to the epoch a subhalo is formed,
  $\mathrm{t}_\mathrm{form}$, versus the lookback time at which the subhalo is
  accreted,  $\mathrm{t}_\mathrm{acc}.$ Different curves show results for different bins of satellite stellar mass, with the gray band showing Poisson errors for the lowest mass sample.
   While
  Fig.~\ref{fig:zstarve_frac} showed the negligible
  contribution of $\zacc$ in our model, this figure demonstrates that satellite quenching in our model is nonetheless connected to accretion time 
 due to the  $\mathrm{t}_\mathrm{form}$--$\mathrm{t}_\mathrm{acc}$ correlation in CDM structure formation.}
\label{fig:tform_tacc}
\end{center}
\end{figure}

The most dramatic result of Figure~\ref{fig:zstarve_frac} is the negligible contribution from $\zacc$, which is
$\lesssim 0.1\%$ at all stellar masses.  While striking, this cannot be directly
interpreted as implying the irrelevance of post-accretion physics on
shaping the colors of satellite galaxies, because the accretion time and formation time of subhalos are correlated.  
This correlation can be directly seen in Fig.~\ref{fig:tform_tacc}, in which we plot the lookback time to the epoch a
subhalo is formed against the lookback time at which it is accreted. We show the mean $t_{\mathrm{form}}$ in bins of $t_{\mathrm{acc}}$ for satellites in several different bins of stellar mass, with 
the gray band illustrating the Poisson error on the mean for the smallest mass sample. Figure~\ref{fig:tform_tacc} demonstrates that the subhalo properties $\zacc$ and $\zform$ are correlated with high statistical significance.\footnote{Note that although the {\em mean} trend is strong, there is roughly $\sim0.5$Gyr of scatter in $t_{\mathrm{form}}$ at fixed $t_{\mathrm{acc}}.$}
Thus, even though neglecting $\zacc$
in our model does not affect our predictions, correlations between satellite quenching and accretion time nonetheless emerge from our model due to 
the connection between the epoch of accretion and the epoch of formation. 
We explore this point in detail in a follow-up paper focusing squarely on satellite quenching in our model.
  
  We have chosen to maintain the exact formalism laid out in Paper I,
  with $\zstarve\equiv
  \mathrm{Max}\left\{\zacc,\zchar,\zform\right\}$.  However, we note
  that in light of the above discussion, we have tested a simple model
  with $\zstarve = \zform$ and find that it performs on par with the
  more complicated model of Paper I. We have also considered
  alternative proxies for halo age beyond $\zform$ that arise often in
  the literature, such as the redshift a halo attains $4\%$ of its
  present day mass, $z_{4\%}$, or the redshift it attains half of its
  present day mass, $z_{1/2}.$  A model based on $z_{4\%}$ works just
  as well as our model, which should not be surprising since the $4\%$
  criterion has already been shown to give a very similar estimate to ours for
    the epoch a halo transitions from the fast- to slow-accretion regime
  \citep{zhao09}.   While a model based on  $z_{1/2}$ correctly
  predicts the general trends of all of the galaxy statistics we
  tested, the results are in significantly worse quantitative
  agreement. This is interesting in light of recent results on {\em
    pseudo-evolution} that imply $z_{1/2}$ is a poorly motivated
  physical proxy for halo age: by the time a typical Milky Way halo
  attains half of its present day mass, its accretion rate is almost
  entirely due to the changing background mean density, to which we
  see no plausible connection to galaxy evolution \citep{diemer_etal13}. Motivated by the prevalence of alternative approaches to the galaxy-halo connection that rely exclusively on host halo mass, we also considered a model that
  rank-ordered on the present day mass of the host halo instead of
  $\zstarve;$ this model made clustering predictions that are grossly
  discrepant with the data. 

The flexibility of the CAM formalism makes exploration of alternative formulations completely straightforward. In the end, $\zstarve$ may
not be the fundamental halo variable that correlates with color; all that can be said from the diverse success of our model is that, whatever this truly fundamental halo property is,
it must correlate strongly with $\zstarve.$ We relegate a more exhaustive exploration of alternative CAM implementations as a task for future work, and comment further 
on this effort in \S~\ref{sec:discussion}.


\section{DISCUSSION}
\label{sec:discussion}

\subsection{Physical Motivation of Age Matching}
\label{subsec:age_motivation}

The physical picture suggested by the myriad successes of abundance
matching is that the depth of a halo's gravitational potential well
($\vmax$) is the single most important quantity in the evolution of a
galaxy. The use of $\zform$ in age matching takes this picture
seriously. The potential well of a halo is primarily built during the
early stage of fast accretion. After the fast-to-slow accretion transition, 
mass accretes onto the halo from the outside-in, and halo mass can 
substantially increase without changing $\vmax$ \citep{wechsler02,zhao03}. 
Even on an individual halo basis, this fact
about structure growth in CDM is readily apparent by visual inspection
of the assembly history of dark matter halos. Thus the use of $\zform$
in age matching is physically motivated by the expectation that
stellar mass build-up slows down once the depth of the central potential is in place, so that 
galaxies with older stellar populations will tend to be found in halos
with potential wells that formed earlier.

\subsection{The Simplicity of CAM}
\label{subsec:age_appeal}

The successes of CAM are particularly interesting in the context of
other approaches to modeling the galaxy-halo connection. First
consider the commonly-used Halo Occupation Distribution \citep[HOD: e.g.,][]{peacock00a, seljak00,
  scoccimarro01a, berlind02, cooray02, zheng07}, or the closely related conditional luminosity function, \citealt[CLF: e.g.,][]{yang03,vdBosch07}. The central quantity in
the HOD is $\PNM,$ the probability that a (host) halo of mass $M$
hosts $N$ galaxies of some type. The HOD formalism was developed over
ten years ago, and there are many noteworthy successes of this
approach to modeling galaxy color and/or star formation rate
\citep[e.g.,][]{skibba_sheth09,zehavi11}.

In one respect, the HOD approach is simpler than ours: in the HOD,
host halo mass is the only halo property that governs galaxy
occupation statistics, whereas CAM models treat halos as objects with
two properties (in the case of age matching, $\vmax$ and
$\zstarve$). On the other hand, HOD models typically require fitting
for a comparatively large number of free parameters in order to model
the color dependence of the spatial distribution of galaxies. As an
example, consider the HOD model explored in \citet{zehavi11}. For
every color-selected luminosity-threshold sample, this model requires
fitting for  five free parameters in a maximum likelihood
analysis. Thus in order to model the clustering at three different
thresholds for both red and blue galaxy samples, this approach
requires fitting for a total of 30 free parameters, since each time
the sample changes, new fits are required.

As another example, consider the HOD model employed in
\citet{tinker_etal13}. Using a standard set of parameters for the HOD
together with parameters describing the stellar-to-halo mass map
$\Mstar(M_{\mathrm{h}}),$ active and passive galaxy populations are
modeled seperately and fit to the observed stellar mass function $\Phi(\Mstar),$
angular clustering $w(\theta),$ and galaxy-galaxy lensing signal
$\Delta\Sigma(R)$. This approach to the galaxy-halo model requires 27
free parameters at each redshift slice.

Semi-analytic models (SAMs) are an alternative to the phenomenological
HOD- and CLF-type modeling. The distinct advantage of the SAM approach
is that these models have the potential to yield constraints on the
actual physical mechanisms responsible for galaxy evolution, since SAM
models attempt to parametrize the litany of baryonic processes that
impact star formation and quenching. However, SAMs typically  require
a large number of finely tuned parameters to describe dynamical
evolution of subhalos and the vast baryonic processes of galaxy
formation
\citep{white_frenk91,kauffmann_etal99,cole_etal94,somerville_primack99,DeLucia_Blaizot07,guo_etal11b}. At
present, there is no consensus in the literature  on the correct SAM
formulation, since quantitative comparisons between different models
are difficult to conduct due to the inherent complexity of this
approach.

By contrast, in our age matching implementation of CAM, we only
require three free parameters: (1) scatter $\sigma$ in the relation
between $\vmax$ and $\Mstar,$ (2) the logarithmic slope $S$ defining
the transition from the fast-to-slow accretion regime of halo growth,
and (3) the characteristic halo mass $M_{\mathrm{char}}$ where star
formation becomes inefficient (recall that $\zacc$ proved
unnecessary). Once these three parameters are fixed, the full
continuous distribution of colors (rather than a binary red/blue
designation) is predicted for {\em all} stellar mass thresholds, since
we implicitly use the color PDF directly from the data. Moreover, the
exact same parameters work equally well for either stellar mass- or
luminosity-threshold samples; as we will show in a companion paper to
this one, the same holds true when using CAM to predict star formation
rates rather than color.  In this sense, our assumption that old halos
host old galaxies results in a dramatically simpler model than
conventional implementations of the HOD.

Though this simplification is tantalizing, it remains to be seen whether the model we have presented in this paper passes a
$\chi^2$ goodness-of-fit criterion for the projected clustering and
galaxy-galaxy lensing signals in SDSS. However, we have not yet fit  for
our three parameters in a maximum likelihood
analysis. We have simply chosen appropriate values from the literature
and kept these fixed throughout this paper. Given how successful our
model has proven to be without varying this small number of parameters, we consider
it plausible that we will be able to achieve a good fit to
the data once we conduct a Monte Carlo Markov Chain exploration of the
CAM parameter space. We consider this effort beyond the scope of the
present work, though we intend to conduct this analysis in future
work, both at low- and high-redshift.

\subsection{Application to Star Formation Measurements}
\label{subsec:sfr}

Though $g-r$ color is generally a good indicator of ongoing star formation in a galaxy, the correspondence is only approximate. 
For instance, star-forming galaxies can often appear red due to the presence of gas and dust \citep{maller_etal09,masters_etal10,Wetzel_Tinker_Conroy12}. 
Additionally, the timescales relevant to, for example, H$\alpha$ indicators of star formation rates (SFR) are significantly shorter than timescales impacting $g-r$ color, 
and so it is reasonable to be skeptical that age matching will fail when connecting present day star formation rate to $\zstarve,$ since this halo property typically occurs in the very distant past.
We show in a companion paper that this is not the case: the SFR predictions of our age matching implementation of CAM are equally successful as the predictions we show here.
Thus a model with just three parameters, with no change to their values, correctly connects dark matter halos to galaxies of a given stellar mass, r-band luminosity, g-band luminosity, and SFR. 

\subsection{Assembly Bias Degeneracies}
\label{subsec:assembly_bias_threat}

One important step in the development of CAM modeling will be to
develop a better understanding of the degeneracies between CAM and the
HOD. Evidently, both classes of models can effectively describe the
color-dependence of the stellar mass function, two-point clustering,
and galaxy-galaxy lensing signal, yet these models are predicated upon
markedly different assumptions. In the HOD, knowledge of halo mass is
sufficient to statistically characterize {\em all} properties of the
galaxies populating a halo; in CAM, galaxy color correlates strongly
with halo age at fixed halo mass; clearly, these are mutually
incompatible assumptions.   We briefly note here that models based
only on host halo mass are unable even in principle  to reproduce {\em
  galactic conformity}: the observation that, at fixed host mass,
halos with a blue central tend to have a bluer satellite
population, and conversely for red centrals and satellites
\citep{weinmann06b}.  This is a natural feature of age matching, since
host halos and subhalos collapse from and co-evolve in the same
over-dense patch of the cosmic density field.  We explore this fully
in a future paper. 

The dependence of galaxy occupation statistics upon some halo property
besides mass generically goes by the name of {\em assembly bias};
\citet{zentner_etal13} recently completed a first step towards
understanding the degeneracies between HOD parameters and the assembly
bias predicted by CAM-type models; in light of the simultaneous
success of these very different approaches to galaxy-halo modeling, we
consider a comprehensive effort to theoretically model and
observationally constrain assembly bias to be necessary in order to
truly understand the connection between galaxies and dark matter
halos.

\subsection{Future Work}
\label{subsec:cam_forward}

Even within the CAM framework there exists degeneracies between
different implementations of the formalism. In \S~\ref{sec:results} we
provide examples of how different definitions of $\zstarve$ can
produce models of comparable success to our fiducial model. For
example, age matching models that either account for, or entirely
ignore, the post-accretion history of satellite galaxies make nearly
identical predictions. Moreover, with Figure~\ref{fig:tform_tacc} we showed that correlations between subhalo age and galaxy color 
can masquerade as intra-host quenching, suggesting that too much emphasis may have been placed on 
the role of post-accretion physics in satellite quenching. We will conduct a more detailed investigation of this point in a follow-up paper dedicated 
to the satellite quenching predictions of age matching.

Additionally, as we showed in Fig.~\ref{fig:sm_color_wp} the influence of $\mchar$ only becomes
significant for stellar masses that contribute minimally to the
clustering in any of our samples. Thus the mass range of our sample
apparently does not require invoking a cutoff in star formation
efficiency, as discussed in \S~\ref{sec:results}. In the interest of continuity with Paper I we
have made no alterations to our definition of $\zstarve,$ though we will return to this issue 
in follow up work in which we explore more exhaustively the variety of different CAM implementations 

The model studied in \citet{masaki13} is in essence a
CAM model in which the conditional abundance is performed with halo
central density rather than $\zstarve.$ It is not particularly
surprising that both models are effective, because in age matching
$\zstarve$ is primarily governed by halo concentration, which is of
course strongly correlated with central density. On the one hand, our
age matching model is physically motivated by the expectation that
older halos tend to host galaxies with older stellar populations,
while we see no plausible, comparably direct physical interpretation
of the central density-based model. On the other hand, the
simultaneous success of $\zstarve$- and central density-based CAM
models implies that caution is required before {\em any} particular
implementation of CAM can be considered to be truly fundamental.

Despite these systematic uncertainties, what is quite clear from our
results is that the CAM formalism provides a framework for the
galaxy-halo connection that makes remarkably accurate predictions for
a rich variety of observational data. We consider the preliminary
success of this framework to be an extremely promising indication that
the cosmic history of star formation in galaxies admits a simple,
elegant theoretical description.



\section{SUMMARY}
\label{sec:summary}

We conclude by summarizing our primary results and conclusions:

\ben
\item [\textbf{1.}] We have made new DR7 SDSS measurements of the 
  projected two-point correlation function, and galaxy-galaxy lensing
  signal, as a function of stellar mass and $g-r$ color. Our measurements appear in Tables 1-6.
  \item [\textbf{2.}] We use these new measurements to test the
  age matching formalism introduced in \citet{HW13a} (Paper I), finding in all cases that our model performs quite well; the level of agreement is particularly remarkable considering the simplicity of the model, and that we have made no alternations to the values of the model parameters used in Paper I. 
\item [\textbf{3.}] We demonstrate a new success of abundance matching: the accurate prediction of the low-redshift galaxy-galaxy lensing signal, and its stellar mass dependence.  
\item [\textbf{4.}] Employing a galaxy group-finder, we showed that our model correctly predicts the colors of central and satellite galaxies, as well as the scaling of these colors with host halo mass.
\item [\textbf{5.}] We present a very general formalism for modeling the galaxy-halo connection called Conditional Abundance Matching (CAM), of which age matching is a specific example. This flexible theoretical framework permits direct investigation of correlations between {\em any} galaxy property and {\em any} halo property.
\item [\textbf{6.}] We make publicly available a mock galaxy catalog constructed from our model at {\tt http://logrus.uchicago.edu/$\sim$aphearin}. 
\een


\begin{table*}
\caption{{\bf SDSS PROJECTED CORRELATION FUNCTION MEASUREMENTS: ALL GALAXIES}. The first column is the the mean radii of galaxies in each logarithmic bin in units of $\Mpc.$  Additional columns show the projected correlation function, $\wprp$, for three stellar mass, volume-limited theshold samples. The diagonal terms of the error covariance matrix are given in the parenthesis.}
\begin{center}
  \begin{tabular}{@{}ccccccc}
\\ \hline \hline
    $r_\mathrm{p}$ & $9.8$ & $10.2$ & $10.6$ \\ \hline
 0.122 & 403.78   (18.80) & 521.56   (25.91) &  646.61   (32.25)\\
 0.173 & 320.76   (15.07) & 405.55   (19.47) & 454.60   (24.62)\\
 0.247 & 250.04   (14.57) & 314.75   (18.65) & 342.78   (19.72)\\
 0.351 & 195.97   (13.40) & 242.73   (16.24) & 260.54   (16.80)\\
 0.500 & 151.47   (11.28) & 184.98   (13.38) & 195.04   (14.68)\\
 0.712 & 116.21    (9.41) & 138.60   (11.13) & 143.84   (12.16)\\
 1.014 &  84.82    (7.57) & 100.96   (9.10) & 104.03    (9.52)\\
 1.445 &  61.24    (5.95) &  71.08   (6.97)  & 74.22    (7.36)\\
 2.058 &  45.65    (5.09) &  52.60    (5.91) &  55.91    (6.11)\\
 2.933 &  35.51    (4.15) &  40.45    (4.80) &  44.06    (4.96)\\
 4.175 &  27.69    (3.49) &  30.78    (3.87) &  32.02    (3.78)\\
 5.942 &  20.58    (2.91) &  22.73    (3.33) &  23.34    (3.28)\\
 8.464 &  14.31    (2.19) &  15.71    (2.46) &  16.90    (2.54)\\
12.048 &   9.67    (1.56) &  10.57    (1.72) &  11.10    (1.83)\\
17.158 &   5.63    (1.28) &   5.96    (1.41) &  6.27    (1.55)\\ \hline

  \end{tabular}

  \label{tab:wp_thresholds_all}
\end{center}
\end{table*}


\begin{table*}
\caption{{\bf SDSS PROJECTED CORRELATION FUNCTION MEASUREMENTS: BLUE GALAXIES}.  Same as Table~\ref{tab:wp_thresholds_all}, but for the blue galaxy samples.}
\begin{center}
  \begin{tabular}{@{}ccccccc}
\\ \hline \hline
    $r_\mathrm{p}$ & $9.8$ & $10.2$ & $10.6$ \\ \hline
 0.121 & 172.54    (9.75) & 207.29   (16.17) & 310.74   (42.11)\\
 0.173 & 146.40    (7.96) & 178.88   (13.34) & 275.17   (28.37)\\
 0.247 & 110.80    (5.77) & 130.81    (8.60) & 178.77   (20.38)\\
 0.352 &  90.95    (5.02) & 108.48    (6.75) & 146.38   (14.16)\\
 0.502 &  69.78    (3.81) &  82.33    (5.20) & 102.14    (9.74)\\
 0.714 &  59.41    (3.21) &  67.92    (4.26) &  81.64    (8.18)\\
 1.017 &  46.87    (2.64) &  54.92    (3.78) &  64.45    (6.39)\\
 1.448 &  37.11    (2.51) &  44.08    (3.47) &  53.11    (5.13)\\
 2.061 &  29.56    (2.45) &  34.58    (3.15) &  38.16    (4.82)\\
 2.935 &  24.16    (2.22) &  28.15    (2.86) &  33.75    (3.63)\\
 4.178 &  19.66    (2.10) &  22.13    (2.57) &  24.72    (3.15)\\
 5.946 &  15.40    (1.98) &  17.57    (2.52) &  19.47    (3.19)\\
 8.466 &  10.85    (1.56) &  12.31    (1.93) &  14.62    (2.42)\\
12.057 &   7.44    (1.23) &   8.21    (1.43) &   8.98    (1.73)\\
17.162 &   4.29    (1.04) &   4.66    (1.18) &   5.52    (1.59)\\ \hline

  \end{tabular}

  \label{tab:wp_thresholds_blue}
\end{center}
\end{table*}


\begin{table*}
\caption{{\bf SDSS PROJECTED CORRELATION FUNCTION MEASUREMENTS: RED GALAXIES}.  Same as Table~\ref{tab:wp_thresholds_all}, but for the red galaxy samples.}
\begin{center}
  \begin{tabular}{@{}ccccccc}
\\ \hline \hline
    $r_\mathrm{p}$ & $9.8$ & $10.2$ & $10.6$ \\ \hline
 0.122 & 403.78   (18.80) & 521.56   (25.91) & 646.61   (32.25)\\
 0.173 & 320.76   (15.07) & 405.55   (19.47) & 454.60   (24.62)\\
 0.247 & 250.04   (14.57) & 314.75   (18.65) & 342.78   (19.72)\\
 0.351 & 195.97   (13.40) & 242.73   (16.24) & 260.54   (16.80)\\
 0.500 & 151.47   (11.28) & 184.98   (13.38) & 195.04   (14.68)\\
 0.712 & 116.21    (9.41) & 138.60   (11.13) & 143.84   (12.16)\\
 1.014 &  84.82    (7.57) & 100.96    (9.10) & 104.03    (9.52)\\
 1.445 &  61.24    (5.95) &  71.08    (6.97) &  74.22    (7.36)\\
 2.058 &  45.65    (5.09) &  52.60    (5.91) &  55.91    (6.11)\\
 2.933 &  35.51    (4.15) &  40.45    (4.80) &  44.06    (4.96)\\
 4.175 &  27.69    (3.49) &  30.78    (3.87) &  32.02    (3.78)\\
 5.942 &  20.58    (2.91) &  22.73    (3.33) &  23.34    (3.28)\\
 8.464 &  14.31    (2.19) &  15.71    (2.46) &  16.90    (2.54)\\
12.048 &   9.67    (1.56) &  10.57    (1.72) &  11.10    (1.83)\\
17.158 &   5.63    (1.28) &   5.96    (1.41) &   6.27    (1.55)\\ \hline

  \end{tabular}

  \label{tab:wp_thresholds_red}
\end{center}
\end{table*}


\begin{table*}
  \label{tab:gg_thresholds_all}
  \caption{{\bf SDSS GALAXY-GALAXY LENSING MEASUREMENTS: ALL GALAXIES}.  The first column is the the mean radii of galaxies in each logarithmic bin, in units of $\kpc.$  Subsequent columns show the galaxy-galaxy lensing signal, $\Delta\Sigma$, in units of $\Msun\mathrm{pc}^{-2}$ for the same three stellar mass, volume-limited threshold samples. Errors in the parenthesis are derived from dividing the survey area into 200 bootstrap subregions and generating 500 bootstrap-resampled datasets.}
  \begin{tabular}{@{}ccccccc}
\\ \hline \hline
    R & $9.8$ & $10.2$ & $10.6$ \\ \hline
  22.31 & 63.70  (21.61) & 74.42  (22.62) & 107.60  (35.44)\\
  27.25 & 30.06  (17.22) & 38.67  (20.61) & 66.02  (30.09)\\
  33.28 & 73.26  (13.96) & 82.86  (16.32) & 93.90  (22.64)\\
  40.65 & 44.88  (12.16) & 53.44  (13.74) & 50.51  (20.24)\\
  49.65 & 28.61   (9.98) & 32.43  (10.95) & 36.85  (15.80)\\
  60.64 & 21.14   (7.67) & 18.12   (8.63) & 31.80  (12.23)\\
  74.07 & 24.35   (5.80) & 19.22   (6.82) & 30.23  (10.61)\\
  90.47 & 15.46   (5.48) & 15.01   (6.16) & 29.77   (8.42)\\
 110.50 &  7.42   (4.60) & 11.58   (5.27) & 18.38   (6.51)\\
 134.96 &  9.16   (3.55) & 11.49   (4.30) & 16.27   (6.25)\\
 164.84 &  6.50   (3.02) &  8.92   (3.48) &  9.30   (4.61)\\
 201.34 &  8.04   (2.28) & 10.52   (2.76) & 18.22   (3.98)\\
 245.92 &  6.10   (1.97) &  6.34   (2.24) &  8.01   (3.47)\\
 300.36 &  5.52   (1.44) &  7.87   (1.75) & 11.24   (2.44)\\
 366.86 &  5.46   (1.32) &  6.33   (1.58) &  7.18   (2.21)\\
 448.09 &  6.39   (1.06) &  6.80   (1.28) &  7.60   (1.90)\\
 547.30 &  4.87   (1.08) &  4.75   (1.26) &  6.17   (1.71)\\
 668.47 &  4.80   (0.81) &  5.65   (0.93) &  5.28   (1.30)\\
 816.47 &  4.30   (0.78) &  5.36   (0.91) &  5.71   (1.19)\\
 997.24 &  3.14   (0.66) &  3.53   (0.74) &  4.78   (0.97)\\
1218.03 &  3.38   (0.62) &  4.15   (0.69) &  4.26   (0.92)\\
1487.70 &  2.45   (0.49) &  2.61   (0.60) &  2.58   (0.70)\\
1817.09 &  2.26   (0.45) &  2.44   (0.54) &  2.21   (0.68)\\
2219.40 &  1.94   (0.38) &  2.20   (0.43) &  1.98   (0.48)\\ \hline
  \end{tabular}
\end{table*}


\begin{table*}
  \label{tab:gg_thresholds_blue}
  \caption{{\bf SDSS GALAXY-GALAXY LENSING MEASUREMENTS: BLUE GALAXIES}.  Same as Table~4, but for the blue galaxy samples only.}
  \begin{tabular}{@{}ccccccc}
\\ \hline \hline
    R & $9.8$ & $10.2$ & $10.6$ \\ \hline
  22.31 &  6.48  (26.63) &   7.84  (33.83) & -13.01  (49.56)\\
  27.25 & 45.09  (24.06) &  58.31  (32.65) &  80.30  (51.75)\\
  33.28 & 63.59  (18.76) &  66.80  (23.73) &  92.62  (40.50)\\
  40.65 & 23.37  (16.81) &  30.48  (20.99) &  28.38  (34.65)\\
  49.65 & 29.98  (13.69) &  28.97  (16.74) &  31.42  (24.16)\\
  60.64 & 14.40   (9.90) &   8.98  (12.68) &  21.85  (18.54)\\
  74.07 & 26.14   (7.77) &  20.19  (10.48) &  30.27  (16.33)\\
  90.47 & 11.48   (7.59) &  14.53  (10.79) &  36.54  (14.80)\\
 110.50 &  4.26   (5.64) &  12.35   (7.42) &  25.49  (11.25)\\
 134.96 &  5.36   (4.89) &   7.00   (6.18) &  14.19   (9.68)\\
 164.84 &  0.17   (4.29) &   3.90   (5.58) &   0.87   (8.74)\\
 201.34 &  2.07   (3.19) &   3.46   (4.27) &  10.81   (6.29)\\
 245.92 &  2.92   (2.97) &   1.68   (3.48) &   1.99   (5.43)\\
 300.36 &  3.14   (2.29) &   4.94   (3.00) &  10.39   (4.27)\\
 366.86 &  2.93   (1.74) &   3.15   (2.21) &   1.52   (3.39)\\
 448.09 &  4.08   (1.48) &   3.25   (2.06) &   4.04   (3.00)\\
 547.30 &  3.03   (1.17) &   2.17   (1.61) &   1.79   (2.42)\\
 668.47 &  2.06   (0.91) &   2.62   (1.22) &   4.11   (1.92)\\
 816.47 &  2.02   (0.88) &   2.69   (1.12) &   3.02   (1.89)\\
 997.24 &  1.26   (0.78) &   1.50   (1.02) &   3.94   (1.41)\\
1218.03 &  1.33   (0.59) &   2.24   (0.81) &   2.91   (1.29)\\
1487.70 &  1.50   (0.51) &   1.91   (0.69) &   2.48   (1.04)\\
1817.09 &  1.23   (0.44) &   1.19   (0.56) &   1.25   (0.89)\\
2219.40 &  0.77   (0.36) &   0.96   (0.40) &   1.15   (0.59)\\ \hline
  \end{tabular}
\end{table*}


\begin{table*}
  \label{tab:gg_thresholds_red}
  \caption{{\bf SDSS GALAXY-GALAXY LENSING MEASUREMENTS: RED GALAXIES}.  Same as Table~4, but for the red galaxy samples only.}
  \begin{tabular}{@{}ccccccc}
\\ \hline \hline
    R & $9.8$ & $10.2$ & $10.6$ \\ \hline
  22.31 & 121.96  (31.69) & 122.83  (33.07) & 177.63  (48.40)\\
  27.25 & 22.89  (27.23) &  31.53  (28.54) &  75.07  (39.61)\\
  33.28 & 99.86  (19.59) & 109.25  (20.74) & 111.89  (28.98)\\
  40.65 & 68.48  (16.84) &  73.14  (17.25) &  68.05  (24.40)\\
  49.65 & 28.77  (14.60) &  37.25  (14.99) &  40.52  (20.49)\\
  60.64 & 29.27  (11.08) &  26.06  (11.38) &  38.25  (17.29)\\
  74.07 & 25.52   (8.76) &  22.61   (9.61) &  36.02  (14.13)\\
  90.47 & 20.70   (7.36) &  15.64   (7.56) &  30.91  (10.31)\\
 110.50 &  8.68   (6.97) &   9.41   (6.94) &  12.21   (9.30)\\
 134.96 & 13.76   (5.82) &  15.03   (6.14) &  17.82   (8.18)\\
 164.84 & 14.43   (4.04) &  12.43   (4.27) &  15.32   (5.47)\\
 201.34 & 16.12   (3.39) &  17.13   (3.62) &  23.43   (5.01)\\
 245.92 &  9.35   (2.61) &   9.46   (2.80) &  11.08   (4.00)\\
 300.36 &  9.97   (2.14) &  11.33   (2.30) &  13.67   (3.22)\\
 366.86 &  8.36   (1.91) &   8.83   (2.00) &  10.65   (2.98)\\
 448.09 &  8.84   (1.66) &   9.18   (1.82) &   9.62   (2.44)\\
 547.30 &  7.07   (1.55) &   6.77   (1.65) &   8.73   (2.22)\\
 668.47 &  7.87   (1.24) &   7.69   (1.30) &   6.02   (1.72)\\
 816.47 &  7.15   (1.07) &   7.51   (1.05) &   7.89   (1.30)\\
 997.24 &  5.27   (0.98) &   5.09   (0.96) &   5.31   (1.16)\\
1218.03 &  5.70   (0.94) &   5.66   (0.94) &   5.17   (1.08)\\
1487.70 &  3.53   (0.79) &   3.19   (0.79) &   2.79   (0.91)\\
1817.09 &  3.33   (0.75) &   3.23   (0.78) &   2.70   (0.86)\\
2219.40 &  3.44   (0.63) &   3.31   (0.62) &   2.65   (0.64)\\ \hline
  \end{tabular}
\end{table*}


\section*{ACKNOWLEDGEMENTS}
\label{sec:ack}

We are particularly grateful to Andrey Kravtsov for insightful 
input at various stages in the development of our model. 
We would like to thank Charlie Conroy, Ramin Skibba, Joel Primack, and Frank van den Bosch for
useful discussions. 
We thank Peter Behroozi for making his halo catalogs and merger trees publicly
available. MRB thanks Peter Behroozi for publicly releasing the ROCKSTAR halo finder under GNU GPLv3, upon which the nearest neighbor finding portion of the code for computing $\Delta\Sigma$ from the Bolshoi simulation was based. 
We would also like to thank John Fahey for {\em Death
  Chants, Breakdowns, and Military Waltzes.}  
  DFW is supported by the National Science Foundation under Award No. AST-1202698. 
  APH supported by the U.S. Department of Energy under contract No. DE-AC02-07CH11359.  
ARZ is supported by the U. S. National Science Foundation through grant AST 1108802 and by the University of Pittsburgh. 
MRB was supported in part by the Kavli Institute for Cosmological Physics at the University of Chicago through grant NSF PHY-1125897 and an endowment from the Kavli Foundation and its founder Fred Kavli.
AAB is supported by NSF grant AST-1109789.
A portion of this work was also supported by the National Science Foundation under grant PHYS-1066293 and the hospitality of the Aspen Center for Physics. 
 This work made extensive use of the NASA Astrophysics Data System and the \verb+arxiv.org+ preprint server.

\bibliography{./darkside2.bib}


\end{document}